\documentclass{aa}  

\usepackage{graphicx}
\usepackage{txfonts}
\usepackage{natbib} 
\bibpunct{(}{)}{;}{a}{}{,} 

\usepackage{verbatim} 
\usepackage{hyperref}   
\hypersetup{colorlinks=true,linkcolor=blue,citecolor=blue,filecolor=blue,urlcolor=blue}

\newcommand{\afe}{[\alpha/{\rm Fe}]}

\begin{document}

   \title{Element abundances of galactic RGB stars in the APO-K2 catalogue} 

   \subtitle{Dissimilarity in the scaling with $\afe$}
  \author{G. Valle \inst{1, 2}, M. Dell'Omodarme \inst{1}, P.G. Prada Moroni
        \inst{1,2}, S. Degl'Innocenti \inst{1,2} 
}
\titlerunning{Chemical investigation on RGB APO-K2 stars}
\authorrunning{Valle, G. et al.}

\institute{
        Dipartimento di Fisica "Enrico Fermi'',
        Universit\`a di Pisa, Largo Pontecorvo 3, I-56127, Pisa, Italy
        \and
        INFN,
        Sezione di Pisa, Largo Pontecorvo 3, I-56127, Pisa, Italy
}

   \offprints{G. Valle, valle@df.unipi.it}

   \date{Received ; accepted }

  \abstract
{}
{We conducted an investigation on the chemical abundances of 4,316 stars in the red giant branch (RGB) phase from  the recently released APO-K2 catalogue. Our aim was to characterize the abundance trends of the single elements with $\afe$, mainly focusing on C, N, and O, which are the most relevant for the estimation of  stellar ages.
}
{
The chemical analysis of the RGB sample involved cross-matching data from the APO-K2 catalogue with individual element abundances from APOGEE DR17.
}
{ 
The analysis detected a statistically significant difference in the [(C+N+O)/Fe] -- $\afe$ trend with respect to the simple $\alpha$-enhancement scenario. This difference remained robust across different choices for the reference solar mixture and potential zero-point calibrations of C and N abundances. The primary discrepancy was a steeper increase in [O/Fe] with $\afe$, reaching a 0.1 dex difference at $\afe = 0.3$. Notably, the impact on the evolutionary timescale of such oxygen over-abundance with respect to the commonly adopted uniform $\alpha$-enhancement is rather limited.  We verified that stellar models computed using an ad hoc O-rich mixture sped up the evolution by only 1\% at $\afe = 0.3$, due to the counterbalancing effects of O enrichment on both the evolutionary timescale and the $Z$-to-[Fe/H] relationship. } 
{} 

   \maketitle

\section{Introduction}\label{sec:intro}

The investigation of the history of formation and evolution of the Galaxy necessitates the establishment of accurate and precise ages for field stars.  In recent years, these Galactic archaeology investigations benefited from a huge amount of high-quality data from different large spectroscopic and astrometric surveys. 
The possibility of obtaining reliable asteroseismic stellar ages \citep[e.g.][]{Miglio2009, Sharma2016, SilvaAguirre2018, Mackereth2020, Warfield2021} coupled with the availability of individual element abundances from APOGEE \citep{Majewski2017}, GALAH \citep{Buder2021}, and LAMOST \citep{Cui2012} 
has opened up new avenues for exploring plausible scenarios of galactic formation \citep[see][for a review]{Matteucci2021}.
These investigations reliably confirmed the presence of a $\afe$ -- [Fe/H] bimodality, already evidenced in previous studies \citep[e.g.][]{Fuhrmann1998, Bensby2003}, and proposed a chronology of the thick and thin disk formation suggesting the presence of both an intermediate-age population at high $\afe$ \citep{Warfield2021}, as well as a young population \citep{Chiappini2015,  SilvaAguirre2018, Grisoni2024}.

Unlike other stellar parameters, stellar ages cannot be directly measured, and the only way to estimate them is by comparison with computed stellar models. While these comparisons, which rely on different assumptions, stellar models, and methods, have been performed in the past for large samples of red giant branch (RGB) stars \citep[e.g.][]{Pinsonneault2014, Martig2015, Warfield2021, Miglio2021}, several aspects of this topic have not been fully explored. It is evident that the age estimates derived from stellar models are as reliable as the assumptions about the physics and chemical input adopted in their computations. 

For instance, the reliability of the effective temperature scale in RGB and its agreement with model predictions is a critical but problematic area.
Two key issues contribute to this problem. First, systematic biases of 50-70 K exist between RGB effective temperature determinations from different surveys \citep[e.g.][]{Hegedus2023, Yu2023}. According to \citet{Warfield2021} such variability would produce a random age uncertainty of about 70\%.
Second, theoretical predictions of the effective temperature for stars in the RGB phase are challenging because of several issues, such as the efficiency of super-adiabatic convection, the outer boundary conditions, and the mass loss rate. As an example, a possible dependence of the mixing-length parameter on metallicity was recently suggested as the cause of the mismatch between theory and observations \citep{Tayar2017}, while \citet{Salaris2018} and \citet{ML} attributed a non-negligible role to the $\alpha$-enhancement.
These significant observational and theoretical uncertainties pose a major challenge because stellar effective temperature is often a critical parameter in grid-based model fitting.

To face this problem, different approaches exist. 
The discrepancy between models and observations might in fact originate from the inaccurate modelling of parameters that primarily influence the effective temperature scale, such as the mixing-length parameter, the boundary conditions, or the relative abundances of Mg and Si \citep{VandenBerg2012}, without significantly impacting the stellar evolutionary timescale.
In this scenario it is possible to estimate the stellar ages without bias as long as the  $T_{\rm eff}$ constraint is not used in the fit. This approach is employed by
 \citet{Martig2015}, \citet{Warfield2021}, and \citet{Warfield2024}, among others, who rely on the mass estimated by scaling relations to interpolate over the stellar models grid.
On  the other hand, 
the discrepancy between the models and observations could stem from other factors, such as variations in the element abundances or an inaccurate scaling of these abundances with $\afe$ \citep[e.g.][]{ML}. There is some evidence in the literature about the different scaling of $\alpha$-elements, in particular regarding a possible oxygen enhancement  with respect to the other elements \citep[e.g.][]{Bensby2005, Nissen2014, deLis2015, Amarsi2019}. In the presence of this disagreement, ages obtained from a comparison with stellar models should be cautiously considered. A variation in the abundances of C, N, O,  which sustain the CNO-cycle,  alters the efficiency of energy production, with a direct impact on the stellar evolutionary timescale  \citep[see e.g.][]{VandenBerg2012, Sun2023, Sun2023b}. Moreover, a variation in the heavy element mixture may alter the $Z$-to-[Fe/H] relation. 
This has significant implications for stellar age calibration, as stellar models rely on both global metallicity $Z$ and initial helium abundance $Y$ as chemical input. Any discrepancy in the $Z$-to-[Fe/H] relation would lead to the use of incorrect stellar tracks for age estimation.

Variations in the initial C, N, and O abundances also affect a relevant chemical tracer of the evolution: the [C/N] abundance ratio. When a star starts ascending the RGB, its convective envelope extends into the interior and brings  material processed from the CNO cycle up
to the surface (known as the first dredge-up). This process leads to a decrease in the [C/N] ratio, and its final value depends on the stellar mass. Therefore, it is not surprising that the [C/N] ratio has been employed to estimate stellar masses and ages for RGB objects  in many works \citep[e.g.][]{Salaris2015b, Martig2016, Hasselquist2019, Casali2019, Miglio2021, Vincenzo2021}. 
However, it is evident that a pipeline, which is calibrated under the assumption of a specific heavy metal abundance ratio and a dredge-up efficiency, and  thus fixing the [C/N]--mass relationship,  will not be able to  predict ages in the presence of both a variation in the element pattern and in the parameters controlling the convective mixing.  

Therefore, a comprehensive examination of the chemical compositions of stars is highly beneficial prior to 
estimating their ages by comparison with models.
An ideal opportunity to conduct this investigation arises from the recent release of the APO-K2 catalogue \citep{Stasik2024}, which contains  high-precision data for 7,673 RGB and red clump stars. The catalogue combines  spectroscopic \citep[APOGEE DR17,][]{Abdurrouf2022}, asteroseismic \citep[K2-GAP,][]{Stello2015}, and astrometric \citep[Gaia EDR3,][]{Gaia2021} data. 
The availability of this information allows  age estimates to be obtained for these stars, extending the investigations performed in the past \citep{Pinsonneault2014, Stello2015, Martig2015, Tayar2017, Salaris2018, Warfield2021}.
In particular, the possibility of using scaling relations to accurately estimate stellar masses allows us  to obtain precise age estimates, which are of obvious importance for Galactic archaeology investigations.

In this paper we match data in the APO-K2 catalogue with single element abundances from APOGEE DR17 \citep{Abdurrouf2022}. We devote particular attention to the analysis of the  C, N, and O abundances; to their variation in different [Fe/H] and $\afe$ ranges; and to the induced differences in the $Z$-to-[Fe/H] relation due to CNO variations. 
This investigation sets the stage for a detailed investigation of possible age bias due to a variation in the chemical abundances pattern in the stellar model computations (Valle et al. in preparation). We present here a preliminary restricted analysis.

\section{Data selection and catalogue cross-matching}\label{sec:methods}

The sample considered in the investigation was selected from the APO-K2 catalogue. Stars entering in the analysis were selected as follows:
\begin{itemize}
        \item Stars in the RGB phase (according to the \texttt{EV} flag in the APO-K2 catalogue), with no missing $\Delta \nu$ and $\nu_{\rm max}$ (5074 stars).
        \item  After applying the initial selection criteria, we further refined the sample by considering only stars with an estimated mass (from scaling relations) within the range [0.75, 1.90] $M_{\sun}$, $T_{\rm eff} > 4000$ K, and no binary flag (4638 stars). This additional filtering eliminated several outliers and unresolved binaries. Stars with mass below 0.75 $M_{\sun}$  were rejected as artefacts because such objects cannot be in the RGB given their evolutionary timescale.
        \item To further refine the sample, we imposed a surface gravity constraint, selecting stars with $\log g \leq 3.25$ (eliminating a few low RGB stars). Additionally, we restricted the sample to medium- to high-precision stars, considering only those with errors in $T_{\rm eff}$ of less than 110 K and in $\afe$ of less than  0.07 dex in the range $\afe \in [-0.15, 0.40]$ (4565 stars).
        \item To clean the sample from possible outliers, the effective temperatures of stars still in the sample were modelled adopting a generalized addictive model approach \citep{Feigelson2012, venables2002modern} using asteroseismic parameters, [Fe/H], and $\afe$ as predictor. Stars with residuals exceeding 120 K were rejected as potential outliers (4381 stars). 
        \item To optimize computational efficiency, we narrowed the scope of our investigation to a metallicity range of $-1.5 \leq {\rm [Fe/H]} \leq 0.4$. The final sample comprises 4316 stars. 
\end{itemize}

Several quantities were extracted from the APO-K2 catalogue: effective temperature, [Fe/H], $\afe$\footnote{Computed as [$\alpha$/M] + [M/H] - [Fe/H].}, $\log g$, $\nu_{\rm max}$, $\Delta \nu$, mass and radius (from scaling relations), and the  classification into three populations (high-$\alpha$, low-$\alpha$, and intermediate-$\alpha$ populations). The ridge line adopted for the separation is shown in Fig.~1 of \citet{Warfield2024}.

Selected stars were cross-matched with chemical information in the APOGEE DR17  dataset, which contains spectroscopic parameters calculated using the APOGEE Stellar Parameters and Chemical Abundances Pipeline (ASCPAP),  completely rewritten for DR17. The primary spectral library uses the SynSpec spectral synthesis code and uses a non local thermodynamic equilibrium treatment for Na, Mg, K, and Ca \citep{Abdurrouf2022}. The simultaneous availability of dynamical and chemical parameters allow a test of the robustness of possible patterns in the chemodynamical space. 
Data selected from APO-K2 catalogue where joined on the APOGEE ID with chemical abundances from the APOGEE DR17 {\it allStars} catalogue. After removing stars with missing abundances,  with the \texttt{STAR\_BAD} or \texttt{CHI2\_BAD} flags set,  3564 objects remain in the dataset. The following abundances were selected: [C/Fe], [N/Fe], [O/Fe], [Na/Fe], [Mg/Fe], [Al/Fe], [Si/Fe], [Ca/Fe], [Ti/Fe]. Starting from these we computed   [(C+N+O)/Fe], [(C+N)/Fe], [C/N],  and [(Mg+Si)/Fe]. Combined element abundances were derived as in the following example:
\begin{eqnarray}
 {\rm [(C+N+O)/Fe]} = \log_{10} ( 10^{{\rm [C/Fe] + C_{\sun} -Fe_{\sun}}} + 10^{{\rm [N/Fe] + N_{\sun} - Fe_{\sun}}} +  \nonumber \\
+  10^{{\rm [O/Fe] + O_{\sun} - Fe_{\sun}}}  ) - \log_{10}(10^{{\rm C_{\sun} -Fe_{\sun}}} + 10^{{\rm N_{\sun} - Fe_{\sun}}} + 10^{{\rm O_{\sun} - Fe_{\sun}}}  ). 
\end{eqnarray}
Here $C_{\sun} =  \log \varepsilon_C = \log(N_C/N_H) + 12 = 8.43$, $N_{\sun} =  7.83$, $O_{\sun} = 8.69$, and $Fe_{\sun} =  7.50$ are the logarithmic abundances from \citet{AGSS09} solar mixture.
It is relevant to note that the  APOGEE DR17 abundances (but not C and N) in the {\it allStars} catalogue are adjusted by applying a zero-point correction to match  those from solar metallicity stars in the solar neighbourhood.

\section{Element abundances}\label{sec:elements}

\begin{figure*}
        \centering
        \includegraphics[width=16.8cm,angle=0]{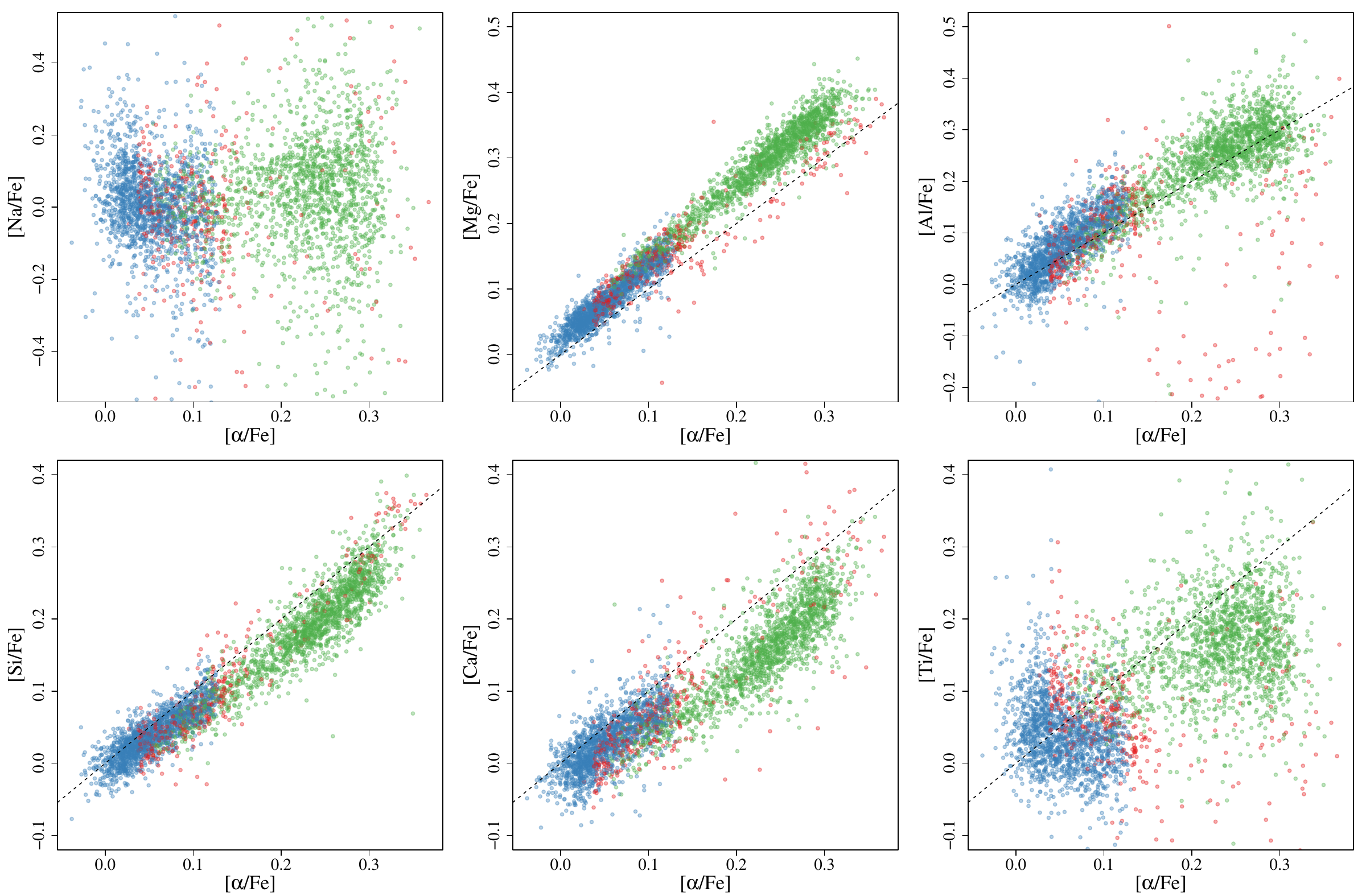}
        \caption{Element abundances vs $\afe$ in the selected RGB sample. The colours in the scatter plots correspond to the groups identified in the APO-K2 catalogue (green = high-$\alpha$ population, blue = low-$\alpha$ population, red = intermediate group). The dashed lines serve as visual aids, highlighting the trend of different elements when assuming the $\alpha$-enhancement from \citet{AGSS09} solar mixture as in Eq.~(\ref{eq:alpha}).
}
        \label{fig:elem-noCNO}
\end{figure*}

\begin{figure*}
        \centering
        \includegraphics[width=16.8cm,angle=0]{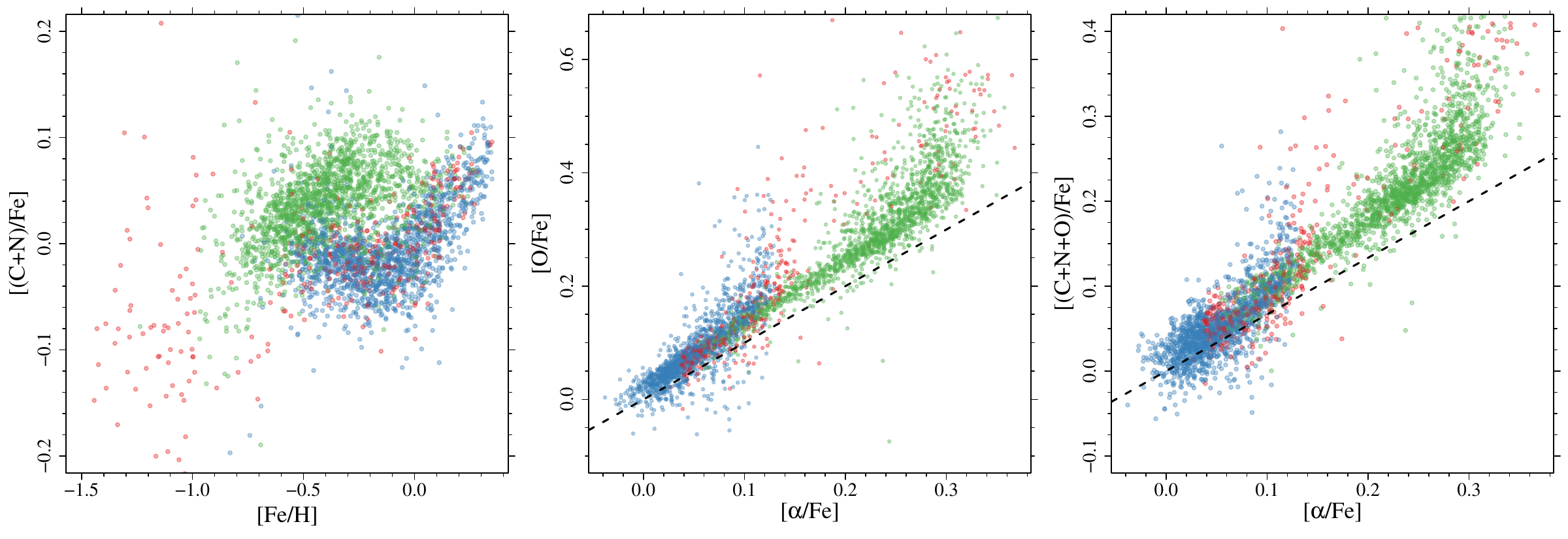}
        \caption{C, N, O abundances in the selected RGB sample. {\it Left}: [(C+N)/Fe] vs [Fe/H]. {\it Centre}: [O/Fe] vs $\afe$. {\it Right}: [(C+N+O)/Fe] vs $\afe$, the dashed lines show the trend when assuming the $\alpha$-enhancement from \citet{AGSS09} solar mixture as in Eq.~(\ref{eq:alpha}). The colour-coding is the same as in Fig.~\ref{fig:elem-noCNO}.}
        \label{fig:elem-CNO}
\end{figure*}

\begin{figure}
        \centering
        \includegraphics[width=8cm,angle=0]{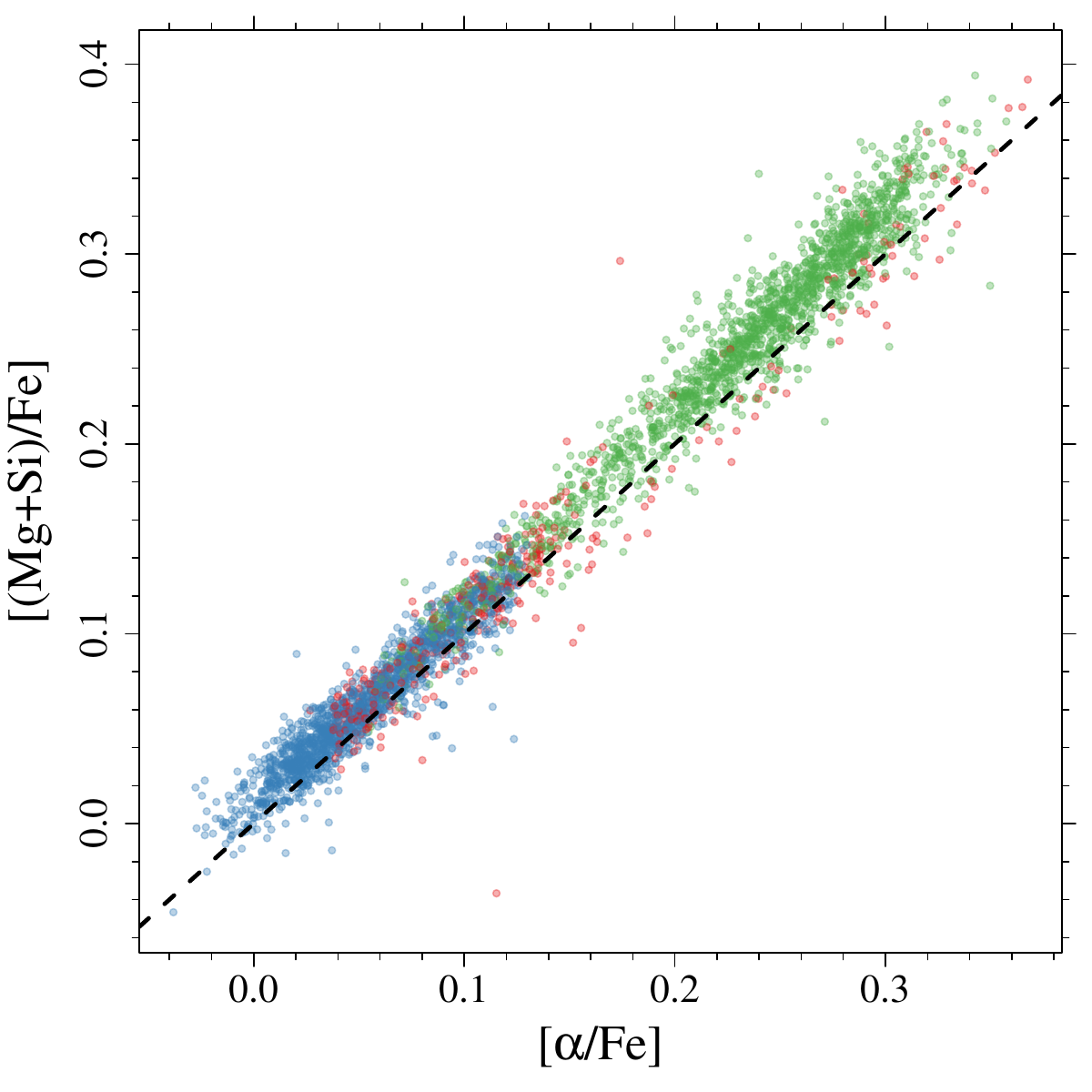}
        \caption{[(Mg+Si)/Fe] vs $\afe$. The dashed line highlights the expected trend of [(Mg+Si)/Fe]]  when assuming the $\alpha$-enhancement from \citet{AGSS09} solar mixture as in Eq.~(\ref{eq:alpha}). The colours code is the same as in Fig.~\ref{fig:elem-noCNO}.}
        \label{fig:MGSIFE}
\end{figure}

Element abundances for the selected RGB stars are presented in Figs.~\ref{fig:elem-noCNO} and \ref{fig:elem-CNO}. Figure~\ref{fig:elem-noCNO} shows the trends of Na, Mg, Al, Si, Ca, and Ti with $\afe$. For the $\alpha$-elements, the radiative opacity tables usually adopted in stellar model computations follow a simple scheme, which assumes the same constant enhancement of the abundances  with respect to the solar value for all the different elements. This results in a linear trend 
\begin{equation}
 {\rm [X/Fe]} = \afe,  \label{eq:alpha}  
\end{equation}
for every element X. The analysis of the  DR17 dataset shows some interesting deviations from this scheme. Magnesium 
exhibits a steeper dependence on
$\afe$ than expected, while the opposite holds for Si. 
Despite these discrepancies, the combined effect of the two trends nearly cancels out, and the evolution of [(Mg+Si)/Fe] with $\afe$ closely follows the expected trend, as shown in Fig.~\ref{fig:MGSIFE}. This result confirms the finding by \citet{Salaris2018}, who reported a similar concordance in the APOKASC sample using the  APOGEE DR13 data release. This close agreement has important consequences for understanding the potential effective temperature mismatch between observations and theoretical predictions. Since the combined Mg and Si abundances are known to establish the RGB effective temperature scale 
\citep{VandenBerg2012, Salaris2018}, 
the agreement of their trend with $\afe$ with what is assumed in Eq.~(\ref{eq:alpha}) excludes variations in these elements as a possible cause for the the $T_{\rm eff}$ discrepancy.
With regard to other elements, Ca exhibits a significant departure from the expected value at high $\afe$, whereas the Al abundance trend is almost the same as
the $\alpha$-elements, in agreement with \citet{Johnson2014, Bensby2017, Duong2019b}. 
The Ti abundance, on the other hand, exhibits greater scatter, as anticipated
because its measurement is considered less reliable than the other  elements \citep{Abdurrouf2022}.
To further explore the reported trends, linear models were fitted for the enhancement of the individual $\alpha$-element [X/Fe] with the global $\afe$. [Ti/Fe] was excluded because of its low reliability. 
The adopted model template was
\begin{equation}
   {\rm [X/Fe]} = \beta_0 + \beta_1 \afe.
\end{equation}
 The results, presented in Table~\ref{tab:ele-alpha}, indicate that the usually adopted assumption that $\beta_1 = 1$, as in Eq.~(\ref{eq:alpha}), might not be accurate. This discrepancy suggests potential limitations in the predictive power of stellar models at high $\afe$ values. A more thorough examination of this discrepancy is beyond the scope of this paper and will be addressed in a specific investigation (Valle et al., in preparation).

\begin{table}
\caption{Linear models for the dependence of the different element  abundances [X/Fe] on $\afe$. }\label{tab:ele-alpha}
\centering
\begin{tabular}{llcc}
  \hline\hline
Element & & Estimate & Std. Error \\ 
  \hline
[O/Fe] & $\beta_0$ & 0.015 & 0.001 \\ 
  & $\beta_1$ & 1.307 & 0.009 \\ 
   \hline
[Mg/Fe] & $\beta_0$ & 0.023 & 0.001 \\ 
  & $\beta_1$ & 1.142 & 0.004 \\ 
\hline 
[Al/Fe] & $\beta_0$  & 0.020 & 0.002 \\ 
 & $\beta_1$ & 0.931 & 0.012 \\ 
   \hline
[Si/Fe] & $\beta_0$  & -0.012 & 0.001 \\ 
 & $\beta_1$ & 0.885 & 0.004 \\ 
   \hline
[Ca/Fe] & $\beta_0$  & -0.018 & 0.001 \\ 
  & $\beta_1$ & 0.784 & 0.006 \\ 
   \hline
\end{tabular}
\end{table}

As a difference from other elements, C and N abundances in the DR17 catalogue are not corrected for zero-point, because their ratio significantly changes during the dredge-up. To better interpret their trends with $\afe$, 
a theoretical investigation of C and N evolution was conducted using a set of stellar models computed employing the FRANEC code \citep{scilla2008} for $\afe = 0.0$ and [Fe/H] = 0.0. Information about the adopted input can be found in \citet{database2012}. Surface abundances of C and N were obtained from these models through linear interpolation in mass, [Fe/H], and $\log g$. Within the explored $\log g$ range, the dredge-up
 already  modified the C and N abundance to their post dredge-up final values. These values depend on several inputs in the stellar evolutionary code, primarily the mixure of  heavy elements  and the efficiency of the superadiabatic convection. To quantify this uncertainty the values of the [C/N] ratio at the end of the dredge-up obtained from the adopted models at various [Fe/H] and stellar masses were compared with those from PARSEC tracks \citep{Nguyen2022}. 
Despite the significant differences between the adopted input of the two stellar codes, including the solar mixture and the mixing-length value, the observed median difference in [C/N] for equivalent models was only 0.04 dex  (PARSEC values being lower). 
This finding provides reassurance about the robustness of the results.
For a detailed investigation of the sources of variability for the post-first dredge-up [C/N] ratio, we refer  to
\citet{Salaris2015b}.

Hence, zero-points for C and N abundances were computed by evaluating the median difference between observed and computed abundances of these elements. A  zero-point of $-0.09$ dex in C and $+0.06$ in N were then applied to observations reported in Fig.~\ref{fig:elem-CNO}. The need for corrections of this magnitude is not surprising because several studies in the  literature, adopting previous APOGEE releases,   report similar results \citep[e.g.][]{Martig2016, Hasselquist2019}.
Furthermore, these corrections are comparable to differences existing among different APOGEE releases. The median change
for C and N between DR14 and DR17 are  $-0.03$ dex and $+0.08$ dex, respectively, while they are $+0.06$ dex and $-0.01$ dex between DR16 and DR17 \citep{Spoo2022}. 

Due to the sensitivity of surface C and N abundances to the assumed dredge-up efficiency, we focus on a tracer that is unaffected by this mixing efficiency, namely [(C+N)/Fe]. This independence is a consequence of the conservation of the total number abundance of C and N during the CNO cycle.
After applying the aforementioned zero-point corrections, the [(C+N)/Fe] abundance around [Fe/H] = 0.0 matches the solar value, as expected. As previously reported in the literature, the median [(C+N)/Fe] abundance in high- and low-$\alpha$ stars are nearly identical (0.03 and 0.0, respectively), with a characteristic increase in both groups with [Fe/H] \citep{Martig2016, Hasselquist2019}, which is attributed to potential AGB enrichment \citep{Masseron2015}.

\begin{figure}
        \centering
        \includegraphics[width=8cm,angle=0]{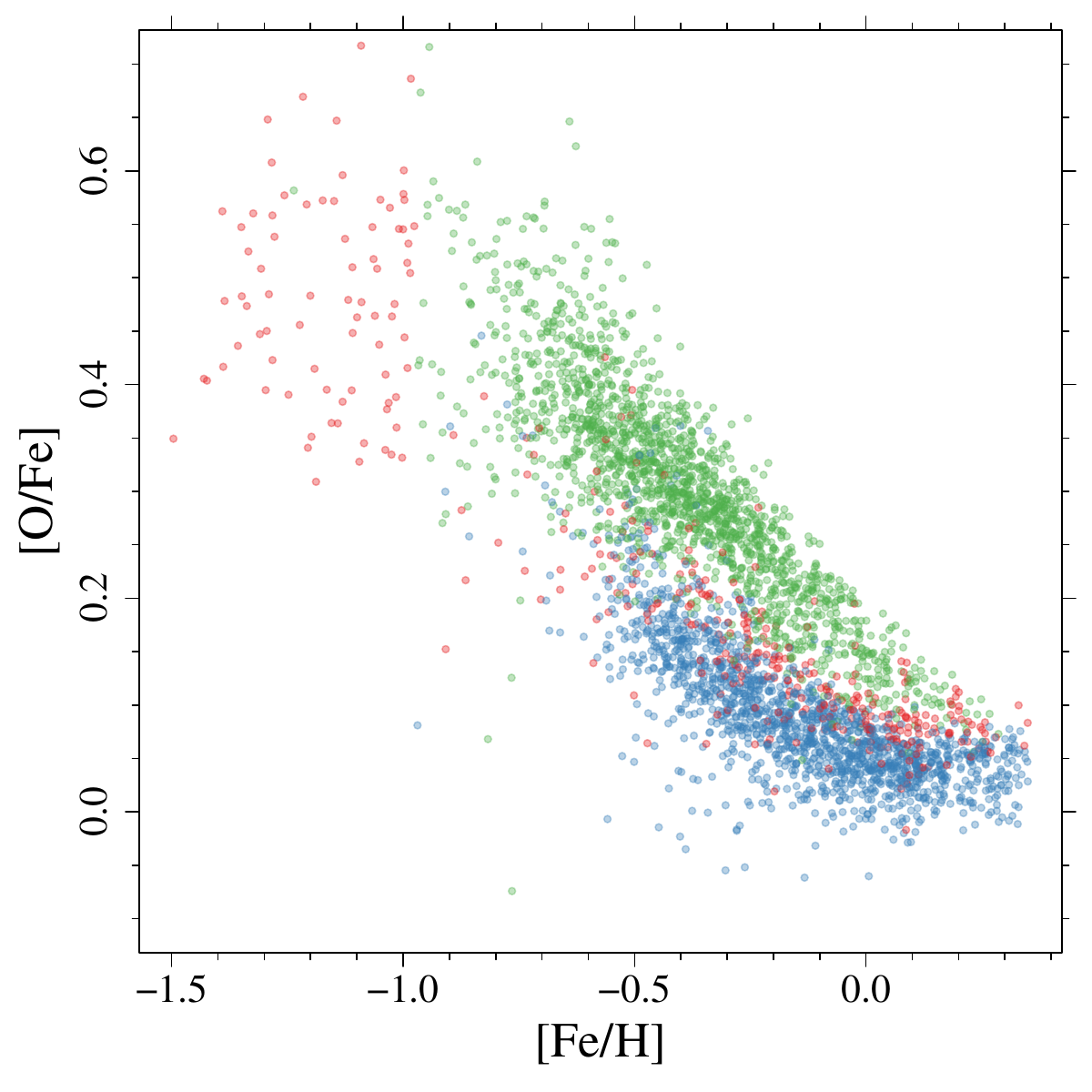}
        \caption{Evolution of [O/Fe] with [Fe/H]. The colour-coding is the same as in Fig.~\ref{fig:elem-noCNO}.}
        \label{fig:Ofeh}
\end{figure}

\begin{figure}
        \centering
        \includegraphics[width=8cm,angle=0]{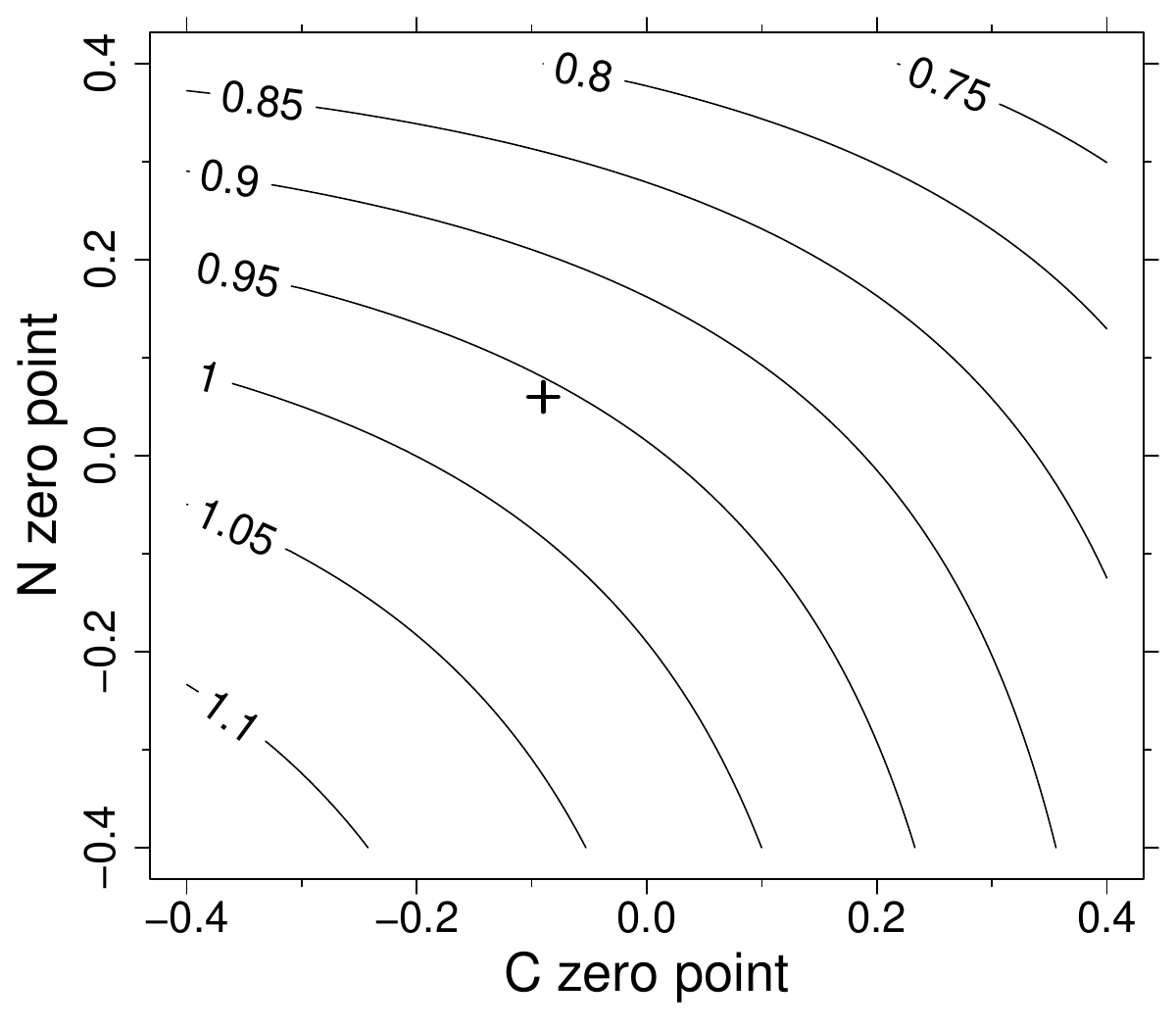}
        \caption{Contour plot showing the value of the [(C+N+O)/Fe] vs $\alpha$ linear fit slope, as a function of different zero-points for carbon and nitrogen abundances. The black cross marks the zero-points adopted in the text.  }
        \label{fig:CNO-zp}
\end{figure}

The central panel in Fig.~\ref{fig:elem-CNO} shows a particularly relevant feature, with interesting consequences in the stellar ages derived from models. The trend of the oxygen with $\afe$ is indeed steeper than expected from Eq.~(\ref{eq:alpha}) (see also Table~\ref{tab:ele-alpha}). The median value of the increase in [O/Fe] around $\afe = 0.20 \pm 0.02$ is 0.25, while it is 0.40 in the range $\afe = 0.30 \pm 0.02$, thus 0.1 dex higher than expected. The finding is not altogether unexpected because 
several studies have proposed that oxygen enhancements may diverge from other $\alpha$-elements \citep[e.g.][]{Bensby2005, Nissen2014, deLis2015, Amarsi2019, Duong2019}.
A distinctive and suspicious increase in [O/Fe] is apparent for both low- and high-$\alpha$ populations at their respective high $\afe$ ends. However, an analysis of the $\chi^2$ values in the APOGEE DR17 dataset does not suggest a problematic fit for these stars, and therefore we retained them in our sample.
This global increase in oxygen abundance adds up to the above-discussed [(C+N)/Fe]. As a result, the trend in the [(C+N+O)/Fe] abundance (right panel in Fig.~\ref{fig:elem-CNO})  deviates markedly from the simple uniform abundance enhancement scaling from the solar \citet{AGSS09}. While the theoretical relation predicts a linear trend with $\afe$ with slope 0.67, a linear fit of the observed abundance (corrected for zero-point offsets) suggests a significantly steeper slope of $0.96 \pm 0.01$.
This conclusion holds, even when adopting a different solar-scaled heavy element mixture as a reference. The \citet{GN93} and \citet{GS98} mixtures both induce a [(C+N+O)/Fe] slope with $\afe$ of 0.70, while  the \citet{Magg2022} mixture produces a slope of 0.64.
However, the inferred slope of the observed trend is sensitive to the adopted zero-point for carbon and nitrogen abundances. 
Therefore, before concluding that the observed trend significantly differs from theoretical expectation, a further check is needed. Figure~\ref{fig:CNO-zp} presents a contour plot of the fitted slope for varying zero-points in C and N abundances. As both zero-points increase, the slope decreases, approaching the theoretical trend for zero-points of around $+0.4$ in both abundances. This scenario is  highly disfavoured, however,  as it would lead to a median  [(C+N)/Fe] value of  0.43 dex in the [Fe/H] solar range [-0.015, 0.015], and a corresponding [(C+N+O)/Fe] of about 0.25. These values significantly exceed the mean values obtained from a sample of B-type stars in the solar neighbourhood by \citet{Nieva2012}, who reported [(C+N)/Fe] = -0.1 dex and [(C+N+O)/Fe] = 0.0 dex. The median [(C+N)/Fe] value would also be higher than those reported by \citet{Masseron2015} and \citet{Hasselquist2019} on stellar samples from earlier APOGEE releases by about 0.4 and 0.35 respectively.

Interestingly, the trend of [O/Fe] versus [Fe/H], shown in Fig.~\ref{fig:Ofeh}, closely matches those reported by \citet{Ramirez2013} and \citet{deLis2015} on quite different samples. In the former paper   oxygen abundances were determined for 825 nearby FGK dwarfs; the stars in that sample were split into thick- and thin-disk sub-populations, according to dynamical considerations. In the latter, 762 solar-like FG stars were divided into thick- and thin-disk stars according to chemical abundances. The resulting trends (Fig.~16 in \citealt{Ramirez2013} and Fig.~13 in \citealt{deLis2015}) are very similar to  Fig.~\ref{fig:Ofeh}. These works did not investigate   the oxygen enhancement with respect to other $\alpha$-elements. The oxygen trend with $\afe$ has recently been investigated by \citet{Sun2023b} in a sample of 67,503 LAMOST and 4,006 GALAH FGK-type dwarf stars. Their findings reveal a general trend of increasing [O/Fe] with $\afe$. However, their analysis is affected by substantial observational scatter, particularly at low $\afe$ values.

\section{Impact on the estimated age. Preliminary analysis}\label{sec:age} 

Figure~\ref{fig:elem-CNO} reveals a higher oxygen-to-iron ratio [O/Fe] than expected from a uniform enhancement of all $\alpha$-elements relative to the solar mixture of \citet{AGSS09}. 
This oxygen over-enhancement reaches   0.1 dex, as [O/Fe]$\approx 0.4$ at $\afe \approx 0.3$. In this section we explore the impact of the impact of  a mixture that is richer in oxygen on derived stellar ages around $\afe = 0.3$.
It is important to note that for this investigation, a standard $\alpha$-enhancement scheme is assumed for all other $\alpha$-elements.
       
The ad hoc radiative opacity tables for  the O-richer mixture were computed employing the OPAL group
\citep{rogers1996} web tools.\footnote{\url{http://opalopacity.llnl.gov/}} 
In our sample, the mean mass of the stars with $\afe \in [0.27,  0.33]$ is $M = 0.97 \pm 0.15$ $M_{\sun}$, and thus we computed stellar models with mass 1.0 $M_{\sun}$. Two different initial metallicities were considered, namely [Fe/H] = $-0.7$ and $-0.5$.  Given the different heavy-element mixtures, the chosen [Fe/H] values correspond to different global metallicities $Z$. 
For the  O-richer mixture, they correspond to $Z = 0.00491$ and 
$Z = 0.00769$, respectively, while for the standard $\afe$= 0.3 with respect to the \cite{AGSS09} solar mixture they correspond $Z = 0.00442$ and $Z = 0.00694$, respectively. The corresponding initial helium abundances ($Y = 0.257$ and 0.261, respectively) are based on a fixed helium-to-metal enrichment ratio $\Delta Y/\Delta Z = 2.0$, and on the primordial helium abundance $Y_p = 0.2471$ \citep{Planck2020}. 
Stellar models were evolved to the point of central hydrogen depletion, conventionally defined as the time when the central hydrogen abundance falls below
$10^{-12}$. 
This age is a proxy for the RGB model's age, given the much shorter timescale of post main-sequence (MS) evolution.

The over-enrichment  of oxygen affects the evolutionary timescale of stellar models in two antagonistic ways. At fixed global metallicity $Z$, an  O-richer mixture  leads to faster evolution \citep{VandenBerg2012, Sun2023, Sun2023b}. On the other hands, at fixed [Fe/H], the O-richer mixture corresponds to higher global metallicity $Z$, which in turn leads to slower evolution. 

The computed stellar models show that the first effect is slightly more important, 
as O-rich models reached the end of the main sequence approximately 0.9\% and 1.2\% faster than the equivalent standard $\afe$=0.3 models. When models are compared at the same $Z$, the discrepancies increase to about $-3.3\%$ and $-4.2\%$, respectively. 
To provide a broader contest for these results, \citet{Sun2023} investigated on the impact of adopting a O-rich mixture in the stellar age estimation for a large sample of MS turn-off and sub-giant stars. They  reported a fractional difference in stellar ages of $-5.3\%$ when adopting a mixture with [O/$\alpha$] = +0.2 from the \citet{GS98} solar mixture. However, a straightforward comparison is probably misleading because of the noticeable differences in the computations. While we present a model-to-model comparison, the results by \citet{Sun2023} involve a fit of stellar observables ($T_{\rm eff}$, luminosity, [Fe/H], $\afe$). Systematic discrepancies between data and models adopted in the fit can therefore play a role that is  difficult to quantify.

This exploratory analysis focused solely on correcting the oxygen mismatch at high $\afe$. By assuming distinct $\afe$-dependent trends for each element, as resulting from observations, we can achieve greater accuracy.
This analysis requires the computation of ad hoc opacity tables and will be presented in detail in a forthcoming paper (Valle et al., in preparation). 

\section{Conclusions}\label{sec:conclusions}

We conducted a chemical investigation using the recently released APO-K2 catalogue \citep{Stasik2024}, encompassing chemical abundances from APOGEE DR17 \citep{Abdurrouf2022}, asteroseismologically determined average quantities from K2-GAP \citep{Stello2015}, and dynamical parameters from Gaia EDR3 \citep{Gaia2021} for approximately 7,500 stars in the RGB and RC stages.
The analysis was restricted to RGB stars. Following a selection procedure aimed at eliminating potential outliers, we obtained a final dataset of 3,564 stars.

The analysis of chemical abundances focused on C, N, and O abundances due to their influence on stellar evolutionary timescales \citep{VandenBerg2012, Sun2023}.
Unlike other elements, APOGEE DR17 carbon and nitrogen abundances are not corrected for zero-points, as dredge-up significantly impacts their surface abundances. Hence, we calibrated them using FRANEC \citep{scilla2008} and PARSEC \citep{Nguyen2022} stellar models, resulting in -0.09 dex for C and +0.06 dex for N. To minimize potential systematics arising from the dredge-up and other mixing processes, we focused on [(C+N)/Fe] and [(C+N+O)/Fe] abundances, which are insensitive to these mixing efficiencies. Following the zero-point corrections, the [(C+N)/Fe] abundance at [Fe/H] = 0.0 aligns with the solar value, as expected. Notably, the trend of oxygen with $\afe$ exhibited a steeper slope than predicted by conventional $\alpha$-enhancement patterns. Within the range $\afe = 0.30 \pm 0.02$, it was 0.1 dex higher than expected. This trend agrees with previous findings, suggesting that the oxygen trend may deviate from other $\alpha$-elements \citep[e.g.][]{Bensby2005, Nissen2014, deLis2015, Amarsi2019}.

As a result of the detected oxygen abundance trend, the global trend in the [(C+N+O)/Fe] versus $\afe$ (slope $0.96 \pm 0.01$) deviates markedly from the theoretical predictions based on scaling from whatever solar mixture. This conclusion is robust for a selection  of solar mixtures \citep{GN93, GS98, AGSS09, Magg2022}, as all these mixtures suggest slopes within 0.64 to 0.70. Since this conclusion is contingent on the adopted zero-point corrections for C and N, we explored the possibility that it could be an artefact. We found that the discrepancy vanishes for zero-point corrections of approximately +0.4 in both abundances. However this scenario is ruled out by observations, as it would result in  too high median [(C+N)/Fe] and [(C+N+O)/Fe] values in the [Fe/H] solar range. In fact, [(C+N)/Fe] would exceed the observed values from \citet{Nieva2012}, \citet{Masseron2015}, and \citet{Hasselquist2019} by about 0.5 dex, 0.4 dex, and 0.35 dex, respectively, while the [(C+N+O)/Fe] abundance would be about 0.25 dex higher than the value reported by \citet{Nieva2012}.

Finally, we investigated the impact of the detected 0.1 dex enhancement in oxygen abundance on stellar age estimates at high $\afe$ values. Unlike previous studies by \citet{Sun2023, Sun2023b}, we relied on a theoretical comparison between stellar models computed using $\alpha$-enhanced mixture scaling from \citet{AGSS09} and a mixture accounting for an additional 0.1 dex enhancement in oxygen abundance. We evolved equivalent $M = 1.0$ $M_{\sun}$ stellar models up to the central hydrogen depletion and quantified the importance of two counteracting factors influencing the evolutionary timescales. While oxygen enrichment accelerates stellar evolution, it also modifies the $Z$-to-[Fe/H] relation, leading to an increase in $Z$ for a given [Fe/H]. This latter effect  offsets the former one, resulting in O-rich models reaching the MS end only approximately 1.0\% faster than the equivalent \citet{AGSS09} $\alpha$-enhanced models. This finding is particularly significant because many recent age estimates for RGB stars \citep[e.g.][]{Martig2015, Warfield2021} do not rely on effective temperature as an observational constraint, as in \citet{Sun2023}, but instead use stellar masses derived from asteroseismic scaling relations. The fact that the estimated ages are relatively insensitive to the potential mismatch in the adopted chemical composition strengthens confidence in the RGB age estimates in the recent literature.

\begin{acknowledgements}
We thank our anonymous referee for the useful comments and suggestions.
G.V., P.G.P.M. and S.D. acknowledge INFN (Iniziativa specifica TAsP) and support from PRIN MIUR2022 Progetto "CHRONOS" (PI: S. Cassisi) finanziato dall'Unione Europea - Next Generation EU.
\end{acknowledgements}

\bibliographystyle{aa}
\bibliography{biblio}

\end{document}